\documentclass[traditabstract]{aa} 

\usepackage{graphicx}

\usepackage{txfonts}
\usepackage{longtable}
\usepackage[authoryear]{natbib}
\bibpunct{(}{)}{;}{a}{}{,}

\def\kms{km~s$^{-1}$}

\begin{document}

\title{Lithium abundance in lower red giant branch stars \\ of Omega Centauri
  \thanks{Based on observations collected at the ESO-VLT under program 096.D-0728.}}

\author{A. Mucciarelli\inst{1,2}, M. Salaris\inst{3}, L. Monaco\inst{4}, P. Bonifacio\inst{5}, 
        X. Fu\inst{1,2}, S. Villanova\inst{6}}
\offprints{A. Mucciarelli}
\institute{
Dipartimento di Fisica e Astronomia, Universit\`a degli Studi di Bologna, Via Gobetti 93/2, I-40129 Bologna, Italy;
\and
INAF - Osservatorio di Astrofisica e Scienza dello Spazio di Bologna, Via Gobetti 93/3, I-40129 Bologna, Italy;
\and
Astrophysics Research Institute, Liverpool John Moores University, 146 Brownlow Hill, Liverpool L3 5RF, United Kingdom 
\and
Departamento de Ciencias Fisicas, Universidad Andres Bello, Fernandez Concha 700, Las Condes, Santiago, Chile
\and
GEPI, Observatoire de Paris, Universit\'{e} PSL, CNRS, Place Jule Janssen 92190, Meudon, France
\and
Universidad de Concepci\'on, Casilla 160-C, Concepci\'on, Chile
}

\authorrunning{A. Mucciarelli et al.}
\titlerunning{Lithium abundance in Omega Centauri}

\abstract{We present Li, Na, Al and Fe abundances of 199 lower red giant branch stars members of the stellar 
system Omega Centauri, using high-resolution spectra acquired with FLAMES at the Very Large Telescope. 
The A(Li) distribution is peaked at A(Li)$\sim$1 dex with a prominent tail toward lower values.  
The peak of the distribution well agrees with the lithium abundances measured in lower red giant branch stars 
in globular clusters and Galactic field stars. 
Stars with A(Li)$\sim$1 dex are found at metallicities lower than [Fe/H]$\sim$--1.3 dex but 
they disappear at higher metallicities. On the other hand, Li-poor stars are found at all the metallicities. 
The most metal-poor stars exhibit a clear Li-Na anticorrelation, with about 30\% 
of the sample with A(Li) lower than $\sim$0.8 dex, while in normal globular clusters 
these stars represent a small fraction. 
Most of the stars with [Fe/H]$>$--1.6 dex are Li-poor and Na-rich.
The Li depletion measured in these stars is not observed in globular clusters with 
similar metallicities and we demonstrate that it is not caused by the proposed 
helium enhancements and/or young ages. Hence, these stars 
formed from a gas already depleted in lithium. 
Finally, we note that Omega Centauri includes all the populations 
(Li-normal/Na-normal, Li-normal/Na-rich and Li-poor/Na-rich stars) observed, 
to a lesser extent, in mono-metallic GCs.}


\maketitle

\section{Introduction}

The emergent and generally accepted picture of stellar populations 
in globular clusters (GCs) is that these stellar systems host different
(chemically distinct) stellar populations. 
In the majority of the cases, the stars of a GC share the same abundances 
of most of the elements, in particular Fe and iron-peak elements
\citep[see e.g.][]{carretta09fe,willman12} demonstrating 
that these systems are not able to retain the ejecta of the supernovae. 
On the other hand, light elements (C, N, O, Na, Mg and Al) exhibit large star-to-star variations and 
often coherent patterns, i.e. C-N, Na-O and Mg-Al anti-correlations 
\citep[see e.g.][]{smith82,kraft92,carretta09g,carretta09u,gratton12,meszaros15,pancino17, bastian17}. 
These chemical patterns are usually interpreted as the signature of
internal pollution by low-energy ejecta of stars
where high-temperature proton-capture cycles (CNO, NeNa and
MgAl cycles) occurred.
The so-called first population (1P) of the cluster is composed by stars with the same 
chemical composition of the original cloud from which the cluster formed. Instead, 
the subsequent second populations (2P) stars show variations of the light element abundances
and they are expected to form from the pristine gas diluted  with the ejecta of the polluter stars. 
A complete and quantitatively successful  model for this self-enrichment process is still lacking, being several 
details still debated \citep[see e.g.][]{renzini15,bastian15}, 
for instance the identification of the main polluter stars. 
Most favorite candidate polluter stars are asymptotic giant branch (AGB) stars
with masses of 4--8 $M_{\odot}$ \citep{ventura01,dercole08,dercole10} and 
fast rotating massive stars \citep{meynet06,prantzos06,decressin07}.

The abundance of lithium -- A(Li)\footnote{A(Li)=$\log{\frac{n(Li)}{n(H)}}+12.00$} -- in GC stars is 
a valuable diagnostic to understand this self-enrichment process, that poses challenges to the
current theoretical models for the formation/evolution of GCs. 
Lithium is one of the few elements created during the Big Bang nucleosynthesis, together 
with H and He. 
In the subsequent evolution of the stars, Li is destroyed in the stellar 
interiors due to proton capture reactions occurring at $\sim2.5\cdot10^6$ K. 
However, Li should be preserved in the stellar envelopes 
of unevolved stars or partially diluted in the photospheres of the stars that have 
undergone processes of mixing. The main mixing episodes are the first dredge-up (FDU) after the 
completion of the main sequence stage, and the extra-mixing episode occurring 
at the luminosity level of the red giant branch (RGB) bump.
Because the proton-capture reactions responsible for the chemical anomalies observed in GC stars 
occurred at temperatures significantly higher ($>10^7$~K) than that of the Li-burning ($\sim10^6$~K), 
the enriched material from which the 2P stars formed should be Li-free.
Therefore, a Li-Na anticorrelation and a Li-O correlation are expected 
within the individual clusters.

Abundances of Li in GCs have been obtained from dwarf and lower RGB (LRGB) stars. 
The latter are stars located between the FDU and the RGB bump. 
Even if measures of A(Li) are limited to a few GCs, the emerging scenario 
turns out to be complex, with some GCs showing evidence of star-to-star scatter in A(Li), 
like in the cases of NGC~6397 \citep{lind09,jonay09}, NGC~6752 \citep{pasquini05,shen10}, 
M4 \citep{monaco12}, NGC~2808 \citep{dorazi15} and 47~Tucanae \citep{dorazi10tuc,dobrov14}. 
Other clusters \citep[M92, NGC~362, NGC~1904, NGC5904 and NGC~6218,][]{boni02,dorazi14,dorazi15} 
show a remarkably homogeneous Li abundance.
Up to now the only clusters that exhibit 
undeniable evidence of correlations between Li and light-elements are NGC~6752 that shows both Li-O correlation 
\citep{shen10} and Li-Na anticorrelation \citep{pasquini05}, NGC~6397 
where 3 Li-poor, Na-rich dwarf stars have been found \citep{lind09} and 
NGC~2808, where some Al-rich RGB stars have A(Li) lower than that in other stars with similar 
Al content \citep{dorazi15}. In M4 a hint of Li-Na anticorrelation is found \citep{monaco12}, 
with a larger A(Li) scatter among 2P stars with respect to 1P stars 
but no evidence of Li-O correlation \citep{m11}. 
47~Tucanae shows the opposite situation, with a statistically significant 
Li-O correlation but not clear evidence of a Li-Na anticorrelation \citep{dobrov14}.

In this paper we investigate the Li content in the stellar system Omega Centauri (NGC~5139) 
using LRGB stars, with the final aim to highlight possible correlations 
with abundances of light elements involved in the chemical anomalies.
Traditionally classified as a GC, Omega Centauri 
exhibits (at variance with the other globulars) a wide iron distribution
\citep{johnson08,johnson09,johnson10,villanova14,pancino11,marino11} 
and a variety of discrete sub-sequences in its colour-magnitude diagram 
\citep[CMD,][]{lee99,pancino00,ferraro04,bedin04,sollima05,sollima07,milone08}, suggesting 
that this system has been able to retain 
the ejecta of the supernovae, experiencing a chemical enrichment history more complex 
than that of a normal GC. According to this evidence,
Omega Centauri is usually interpreted as the remnant of a disrupted dwarf spheroidal 
galaxy \citep{bekki03}. 
On the other hand, its dominant (metal-poor) population shows a clear Na-O anticorrelation 
\citep{johnson10,marino11}, demonstrating that the same self-enrichment process 
observed in mono-metallic GCs occurred also in this stellar system.

The only determination of A(Li) in Omega Centauri stars has been presented by
\citet{monaco10} who analysed a sample of 91 dwarf stars
in the iron range [Fe/H]=--2.0/--1.4 dex, finding an average lithium A(Li)=~2.19 dex 
($\sigma$=~0.14 dex). This abundance is compatible 
with the values usually measured among the Galactic dwarf stars. Measures for light element 
abundances are not available for this sample of dwarf stars.
Because of the faintness of the dwarf stars in Omega Centauri (V$>$18), 
large samples of high quality spectra can be obtained only with a huge effort in terms 
of telescope time. The observation of LRGB stars (instead of dwarf stars) allows to 
study the lithium content in Omega Centauri using large sample of high signal-to-noise (SNR), 
high-resolution spectra, coupling these abundances with those of the light elements 
in order to study the lithium content in 1P and 2P stars.

The paper is structured as follows: Section~2 describes the observations and 
selection of the member stars of Omega Centauri; Section~3 describes the adopted procedure 
for the chemical analysis; the results are presented 
in Section~4, 5 and 6 and discussed in Section 7.

\section{Observations and membership}

%%%%%%%%%%%%%%%%%%%%%%%%%%%%%%%%%%%%%  CMD

\begin{figure}
\includegraphics[width=\columnwidth]{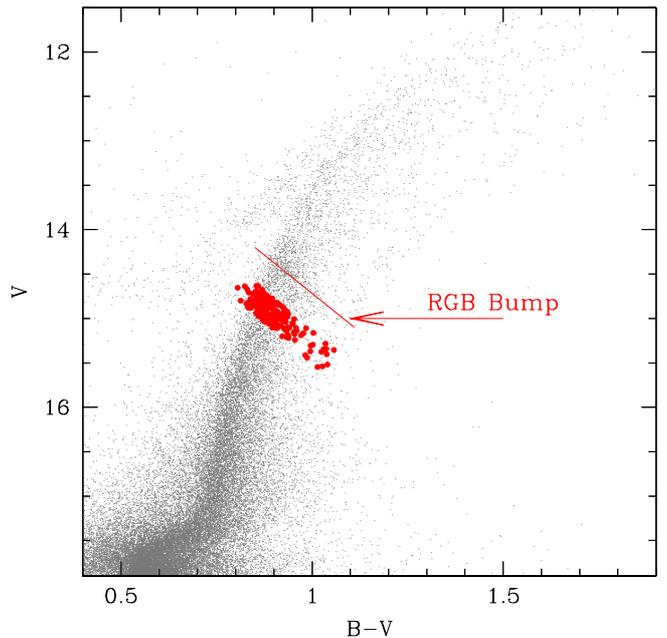}
\caption{CMD of Omega Centauri 
\citep[][stars selected within 750 arcsec from the cluster center]{bellini09}, with 
marked the position of the spectroscopic targets (red circles) and the mean locus (red line) 
of the RGB bump of the different stellar populations of Omega Centauri.}
\label{cmd}
\end{figure}

The observations have been collected under the ESO program 096.D-0728 (PI: Mucciarelli) 
using the multi-object spectrograph FLAMES \citep{pasquini00} at the Very Large Telescope. 
The GIRAFFE+UVES combined mode has been used, allowing the simultaneous allocation of 132 
mid-resolution GIRAFFE fibers and 8 high-resolution UVES \citep{dekker00} fibers.

%%%% set-ups
All the stars have been observed with three FLAMES-GIRAFFE set-ups, namely 
HR12 ($\Delta\lambda$=5821-6146 \AA\ , R$\sim$20000),
HR13 ($\Delta\lambda$=6120-6405 \AA\ , R$\sim$26000) and
HR15N ($\Delta\lambda$=6470-6790 \AA\ , R$\sim$19000),
allowing the measure of several Fe spectral lines, as well as 
the Li line at 6708 \AA , the Na~D doublet at 5890-5896 \AA\  and the Al~I 
doublet at 6696-6698 \AA\ . 
Two observations of 1350 s each have been secured for HR12, 
4 exposures of 1800 s each for HR12 and 6 exposures of 2700 s each for HR15N. 
Two target configurations have been taken, observing  a total of 211 stars.

%%%% selection
The targets have been selected from the \citet{bellini09} BVRI WFI\@2.2m photometric
catalogue, picking stars $\sim$0.6 mag fainter than the mean locus of the RGB bump. 
These stars are hence located after the completion of the FDU and before the onset 
of the extra-mixing usually observed after the RGB bump.
Fig.~\ref{cmd} shows the positions of the observed targets (red circles), 
together with the mean locus of the RGB bump (red thick line). Only stars without 
close companions of comparable or brighter luminosity 
have been selected, in order to avoid stellar contamination within the fibers.

%Spectral reduction
All the spectra have been reduced with the standard ESO GIRAFFE pipeline\footnote{http://www.eso.org/sci/software/pipelines/}, 
including bias-subtraction, flat-fielding, wavelength calibration with a reference Th-Ar lamp 
and extraction of the 1-dimensional spectra.

%RV
Radial velocities (RVs) have been measured for each individual spectrum 
from the position of several metallic lines using the code {\tt DAOSPEC} \citep{stetson08}.
After the correction for the corresponding heliocentric velocity, the spectra 
of each star have been coadded together and used for the chemical analysis. 
Typical SNR per pixel are 70-100 for HR12, 120-150 for HR13 and
180-230 for HR15N.

We excluded from the following chemical analysis 6 stars with RV not compatible 
with that of Omega Centauri and 4 stars with RV dispersion larger than 1.7 km/s 
(while all the other targets have RV dispersions significantly smaller than 1 km/s) 
and considered as candidate binary stars. 
Finally, we derived chemical abundances for a total of 201 member stars of Omega Centauri. 
All the main information (coordinates, magnitudes, RVs) of the member stars 
are listed in Table~\ref{info}.

\begin{table*}
\begin{center}
\caption{Main information on the member LRGB targets of Omega Centauri. 
Coordinates and B, V, I magnitudes are from \citet{bellini09}, $K_{s}$ magnitudes 
are 2MASS database \citep{skrutskie}.
This table is available in its entirety in machine-readable form.}
\label{info}
\begin{tabular}{lccccccc}
\hline
ID &   RA      &   Dec     & B & V & I & $K_{s}$ & ${\rm RV_{hel}}$   \\
             & (J2000)   &  (J2000)  &   &   &   &   &  (\kms)     \\
\hline   
 22580      &	 201.723137 &	--47.661730 &  15.630  &  14.747  & 13.660 & 12.293 & +216.2$\pm$0.2	\\  
 25939      &	 201.832739 &	--47.653890 &  15.818  &  14.950  & 13.874 & 12.394 & +227.8$\pm$0.1	\\  
 31135      &	 201.784629 &	--47.643055 &  15.707  &  14.843  & 13.760 & 12.370 & +239.0$\pm$0.1	\\  
 35536      &	 201.750357 &	--47.634749 &  15.634  &  14.763  & 13.677 & 12.269 & +224.8$\pm$0.1	\\  
 45849      &	 201.765455 &	--47.617715 &  15.616  &  14.762  & 13.752 & 12.461 & +252.3$\pm$0.1	\\  
 49844      &	 201.735347 &	--47.611785 &  15.890  &  14.981  & 13.868 & 12.450 & +223.9$\pm$0.1	\\  
 56555      &	 201.683982 &	--47.602406 &  16.002  &  15.053  & 13.948 & 12.548 & +218.5$\pm$0.1	\\  
 57780      &	 201.709221 &	--47.600836 &  15.972  &  15.041  & 13.951 & 12.592 & +224.0$\pm$0.2	\\  
 64366      &	 201.683249 &	--47.592755 &  15.904  &  15.013  & 13.914 & 12.460 & +224.6$\pm$0.2	\\  
 67162      &	 201.724149 &	--47.589388 &  15.880  &  14.976  & 13.897 & 12.504 & +245.2$\pm$0.1	\\  
 70485      &	 201.812966 &	--47.585439 &  15.568  &  14.723  & 13.644 & 12.303 & +237.4$\pm$0.1	\\  
 71476      &	 201.601550 &	--47.584175 &  15.551  &  14.699  & 13.641 & 12.246 & +227.1$\pm$0.2	\\  
 73028      &	 201.820008 &	--47.582544 &  15.665  &  14.787  & 13.706 & 12.339 & +233.3$\pm$0.1	\\  
 73743      &	 201.920362 &	--47.581721 &  15.851  &  14.985  & 13.907 & 12.533 & +253.1$\pm$0.1	\\  
 73986      &	 201.741336 &	--47.581473 &  15.683  &  14.787  & 13.692 & 12.197 & +259.6$\pm$0.1	\\  
 74878      &	 201.892105 &	--47.580466 &  15.712  &  14.846  & 13.764 & 12.419 & +243.7$\pm$0.1	\\  
 77093      &	 201.563334 &	--47.578011 &  15.792  &  14.919  & 13.832 & 12.487 & +222.1$\pm$0.2	\\  
\hline
\end{tabular} 
\end{center}
\end{table*}

\section{Chemical analysis}

%%% Atmospheric parameters
Effective temperatures (${\rm T_{eff}}$) and surface gravities (log~g) have been derived from 
the photometric information. 
${\rm T_{eff}}$ have been obtained by averaging ${\rm T_{eff}}$ derived from $(B-V)_0$, $(V-I)_0$ and $(V-K_{s})_0$  
from the \citet{alonso99} transformations. B, V and I magnitudes are from \citet{bellini09}, 
$K_s$ magnitude from 2MASS database \citep{skrutskie}. 
Because the \citet{alonso99} calibration 
adopted the Johnson I-filter, the Cousin I-band magnitudes by \citet{bellini09} 
have been transformed to the Johnson photometric system adopting the transformation provided by 
\citet{bessell79}.
The $K_s$ 2MASS magnitudes have been transformed to the Telescopio Carlos Sanchez 
photometric system \citep[used by][]{alonso99} by means 
of the relations by \citet{carpenter} and \citet{alonso98}.

The dereddened colours have been obtained by adopting 
the extinction coefficients from \citet{mccall04} and 
the colour excess quoted in the \citet{harris10} catalogue 
(E(B-V)=~0.12 mag). Note that Omega Centauri is 
not significantly affected by differential reddening \citep{bellini09} 
and this effect has been neglected in the atmospheric parameters determination,

2MASS $K_{s}$ magnitudes are available for 143 stars and for them we averaged the three 
${\rm T_{eff}}$ values. For 54 stars we used only ${\rm T_{eff}}$ derived from $(B-V)_0$ and $(V-I)_0$, 
and for 4 stars (for which I and $K_{s}$ magnitudes are not available) ${\rm T_{eff}}$ from $(B-V)_0$ have been 
adopted. We checked that no significant offset exists among the different ${\rm T_{eff}}$  scales: 
the average differences between the ${\rm T_{eff}}$  scales are
${\rm T_{eff}^{(V-K_{s})}}-{\rm T_{eff}^{(B-V)}}$=+49 K ($\sigma$=+74 K),
${\rm T_{eff}^{(V-K_{s})}}-{\rm T_{eff}^{(V-I)}}$=+45 K ($\sigma$=+51 K) and
${\rm T_{eff}^{(B-V}}-{\rm T_{eff}^{(V-I)}}$=--30 K ($\sigma$=+61 K).

%%% gravity
Gravities have been derived through the Stefan-Boltzmann relation, adopting 
the photometric ${\rm T_{eff}}$ described above, a true distance modulus of $(m-M)_0$=13.70 mag
\citep{bellazzini04}, the bolometric corrections calculated according to \citet{alonso99}, while
the evolutive mass have been estimated according to a grid of theoretical isochrones 
of different metallicities from the BaSTI database \citep{pietr06}.

%%% vt
Microturbulent velocities ($v_t$) have been derived spectroscopically by erasing any trend 
between the line strength and the iron abundance using $\sim$35-40 Fe~I lines.

% Fe, Na, Al
Chemical abundances for Fe and Al have been derived with the code {\tt GALA} 
\citep{m13g}\footnote{http://www.cosmic-lab.eu/gala/gala.php} 
by comparing the theoretical and measured equivalent widths (EWs) 
of unblended metallic lines. The adopted model atmospheres have been 
calculated with the last version of the code 
{\tt ATLAS9}\footnote{http://wwwuser.oats.inaf.it/castelli/sources/atlas9codes.html} \citep[see][]{sbordone04,kurucz05}.
EWs have been measured with the code {\tt DAOSPEC} \citep{stetson08} 
managed through the wrapper {\tt 4DAO} \citep{4dao}\footnote{http://www.cosmic-lab.eu/4dao/4dao.php}. 
For some stars, Al lines are too weak to be measured and we computed upper limits 
adopting as EW 3 times the uncertainty calculated according to the \citet{cayrel88} formula.

% Li 
Abundances for Li and Na have been derived through a $\chi^2$-minimization, 
performed with our own code {\tt SALVADOR}, between the observed and 
synthetic spectra. The synthetic spectra have been computed with the 
code {\tt SYNTHE} \citep{sbordone04, kurucz05}, including all the atomic and molecular 
transitions from the Kurucz/Castelli linelist.
The Li abundances have been derived from the Li resonance doublet at $\sim$6708 \AA\ , 
including the corrections for non-local thermodynamical equilibrium from \citet{lind08}. 
Lithium abundances have been derived for 168 targets, while for 33 stars the Li line is 
too weak and only upper limits are provided.
Na abundances from the Na D doublet at 5890--5896 \AA\ 
have been corrected for departures from local thermodynamical equilibrium 
adopting the corrections computed by \citet{lind11}.

Abundance uncertainties have been calculated by adding in quadrature the 
errors arising from the measurement procedure (EW or spectral fitting) and those arising from 
the adopted atmospheric parameters (uncertainties in ${\rm T_{eff}}$ and log~g have 
been added directly because the two parameters are correlated). For Fe and Al, we consider as internal error the 
dispersion normalized to the root mean square of the number of used lines.
For Li and Na abundances, derived from spectral synthesis, we estimated the uncertainties 
in the fitting procedure resorting to MonteCarlo simulations. 
For each star, a sample of 300 noisy spectra has been created, injecting Poissonian noise 
in the best-fit synthetic spectrum (rebinning to the FLAMES-GIRAFFE pixel-size). 
Each sample of spectra has been analysed with the same procedure 
adopted for the real spectra. The dispersion of the derived abundance distribution 
has been assumed as 1$\sigma$ uncertainty in the abundance.
Uncertainties due to the atmospheric parameters have been derived by repeating the analysis 
varying each time one only parameter by the corresponding uncertainty 
($\sigma_{\rm T_{eff}}=\pm50 K$, $\sigma_{\rm v_{t}}=\pm0.1$ \kms, 
$\sigma_{logg}=\pm0.1$).

Typical formal uncertainties in A(Li) are of the order of 0.05-0.07 dex, being the uncertainty 
in ${\rm T_{eff}}$ the dominant source of error for these abundances, while 
the error due to the fitting procedure is smaller than 0.03 dex, because of the high SNR of the spectra.
Atmospheric parameters, abundance ratios and their uncertainties for all the member stars are listed 
in Table~\ref{info2}.

\section{[Fe/H] distribution}
\begin{figure}
\includegraphics[width=\columnwidth]{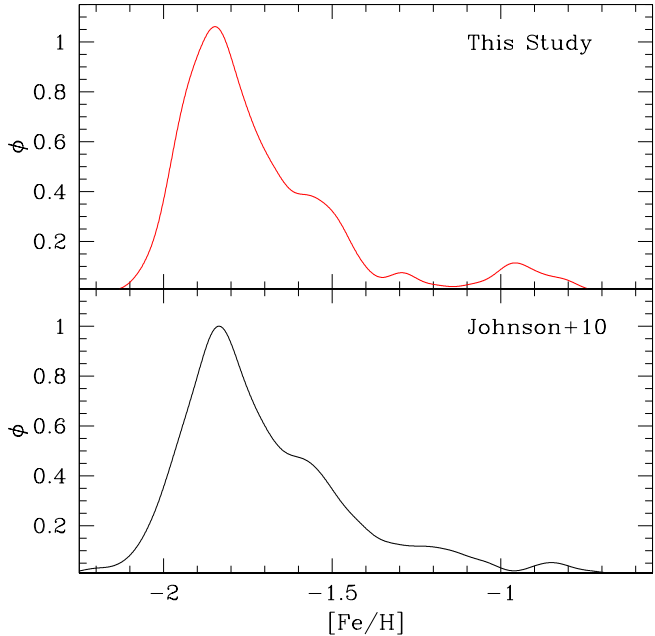}
\caption{Generalized histograms for the LRGB stars of Omega Centauri 
discussed in this study (upper panel) and for the bright RGB stars discussed 
by \citet{johnson10}, shifted by --0.08 dex in order to match the main peaks 
of the two distributions.}
\label{irondis}
\end{figure}

Fig.~\ref{irondis} shows  the [Fe/H] distribution of the entire sample (upper panel) as a generalized histogram 
\citep[a representation that removes the effects due to the choice of the starting 
point and of the bin size, by taking into account the uncertainties in each individual value, see ][]{laird88}.

%%%%%%%%%%%%%%%%%%%%%%%%%%%%%%%%%%%%% Fe distribution

The iron distribution ranges from --2.06 dex to --0.76 dex, 
with a main peak at [Fe/H]$\sim$--1.85 dex, a clear second peak at [Fe/H]$\sim$--1.55 dex 
and a long metal-rich tail with a clump of stars at [Fe/H]$\sim$--0.9 dex 
and a possible small peak at [Fe/H]$\sim$--1.3 dex.
This distribution is qualitatively similar to those previously discussed in other 
studies \citep{johnson08,johnson09,johnson10,marino11}. 
The lower panel of Fig.~\ref{irondis} shows the [Fe/H] distribution 
by \citet{johnson10} that presented the largest (855 stars) dataset of abundances for RGB stars of 
Omega Centauri. We note that the main peak of our distribution is slightly most
metal-poor than that by \citet{johnson10}. 
We estimate a difference of --0.08 dex between 
the median values of the metal-poor components of the two distributions. 
In Fig.~\ref{irondis} the [Fe/H] distribution by \citet{johnson10} 
has been shifted by --0.08 dex in order to match the main peaks of the two distributions. 
Because there are no stars in common with this study (that is focussed 
on RGB stars brighter than the RGB bump) we cannot directly investigate the origin of this small 
difference. However, we noted that the V-band magnitudes used by \citet{johnson10} are 
brighter than those of \citet{bellini09} by about 0.1 mag (while the $K_s$-band magnitudes 
are the same), leading to slightly hotter $T_{\rm eff}$ with respect to our ones.
Despite this small offset, the two distributions are very similar.

Following the scheme used by \citet{johnson10} to describe their [Fe/H] distribution, 
we adopted the nomenclature proposed by \citet{sollima05} to associate the individual 
RGBs observed in the CMD of Omega Centauri with the components of its iron distribution. 
The main peak in our [Fe/H] distribution can be associated to the RGB-MP (the main giant branch 
observed in the CMD of Omega Centauri), 
the second peak to the RGB-Int1, while the most metal-rich peak to the anomalous RGB. 
\citet{sollima05} identified other two RGBs, with metallicites intermediate between those of RGB-Int1 
and the anomalous RGB. \citet{johnson10} identified the stars in the range 
[Fe/H]$\sim$--1.4/--1.0 dex in their [Fe/H] distribution as belonging to 
the RGB-Int2 and RGB-Int3.
In our  distribution this [Fe/H] range is less populated with respect 
to the \citet{johnson10} distribution, probably because of a poor sampling of the 
reddest stars (our targets are confined to a narrow strip on the CMD, see Fig.~\ref{cmd}).

\begin{table*}
\begin{center}
\caption{Atmospheric parameters and abundance ratios (and corresponding uncertainties) 
for the member LRGB stars in Omega Centauri. 
This table is available in its entirety in machine-readable form.}
\label{info2}
\begin{tabular}{lccccccc}
\hline
ID &   ${\rm T_{eff}}$     &  log~g     & $v_t$   & [Fe/H] & ${\rm A(Li)_{NLTE}}$ & ${\rm [Na/Fe]_{NLTE}}$ & [Al/Fe]   \\
   &    (K)          &            &  (\kms) &        &   &   &       \\
\hline   
 22580      &  4901   &  2.31  &  1.5  &  --1.98$\pm$0.07   & 1.02$\pm$0.05   &--0.42$\pm$0.08 & $<$0.94  \\  
 25939      &  4890   &  2.39  &  1.6  &  --1.71$\pm$0.07   & 1.00$\pm$0.05   &--0.26$\pm$0.07 & $<$0.45  \\  
 31135      &  4913   &  2.36  &  1.3  &  --1.78$\pm$0.07   & 0.85$\pm$0.05   & +0.18$\pm$0.06 & $<$0.57  \\  
 35536      &  4907   &  2.33  &  1.5  &  --1.60$\pm$0.07   & 0.71$\pm$0.06   & +0.31$\pm$0.07 &       +1.29$\pm$0.08  \\  
 45849      &  5068   &  2.41  &  1.6  &  --1.49$\pm$0.07   & $<$0.67	      & +0.16$\pm$0.08 & $<$0.27  \\  
 49844      &  4848   &  2.38  &  1.7  &  --1.67$\pm$0.07   & 1.03$\pm$0.07   &--0.25$\pm$0.07 & $<$0.49  \\  
 56555      &  4843   &  2.41  &  1.6  &  --1.65$\pm$0.07   & 0.90$\pm$0.05   & +0.20$\pm$0.05 &       +0.51$\pm$0.10  \\  
 57780      &  4879   &  2.41  &  1.3  &  --1.87$\pm$0.07   & 0.91$\pm$0.05   &--0.15$\pm$0.07 & $<$0.77  \\  
 64366      &  4857   &  2.41  &  1.4  &  --1.77$\pm$0.07   & 0.90$\pm$0.06   &--0.20$\pm$0.05 & $<$0.50  \\  
 67162      &  4899   &  2.41  &  1.4  &  --1.54$\pm$0.07   & 0.78$\pm$0.06   & +0.39$\pm$0.05 &       +1.39$\pm$0.07  \\  
 70485      &  4943   &  2.32  &  1.5  &  --1.86$\pm$0.07   & 0.84$\pm$0.05   & +0.13$\pm$0.07 &       +1.09$\pm$0.08  \\  
 71476      &  4944   &  2.31  &  1.6  &  --1.80$\pm$0.07   & 0.82$\pm$0.05   & +0.09$\pm$0.07 &       +1.10$\pm$0.07  \\  
 73028      &  4930   &  2.35  &  1.5  &  --1.49$\pm$0.07   & 0.72$\pm$0.05   & +0.39$\pm$0.05 &       +1.37$\pm$0.05  \\  
 73743      &  4921   &  2.41  &  1.3  &  --1.84$\pm$0.07   & 1.08$\pm$0.06   & +0.09$\pm$0.05 & $<$0.72  \\  
 73986      &  4852   &  2.31  &  1.4  &  --1.58$\pm$0.07   & 0.81$\pm$0.05   & +0.24$\pm$0.06 &       +1.43$\pm$0.08  \\  
 74878      &  4927   &  2.36  &  1.5  &  --1.81$\pm$0.07   & 1.02$\pm$0.06   & +0.07$\pm$0.07 & $<$0.59  \\  
 77093      &  4918   &  2.38  &  1.4  &  --1.85$\pm$0.09   & 1.04$\pm$0.07   &--0.29$\pm$0.09 & $<$0.77  \\  
\hline
\end{tabular} 
\end{center}
\end{table*}

\section{Lithium in Omega Centauri}

We discuss here the A(Li) distribution in 199 LRGB stars of Omega Centauri, excluding 
two targets that reveal a high Li content and that 
will be discussed in a forthcoming paper.

\begin{figure}
\includegraphics[width=\columnwidth]{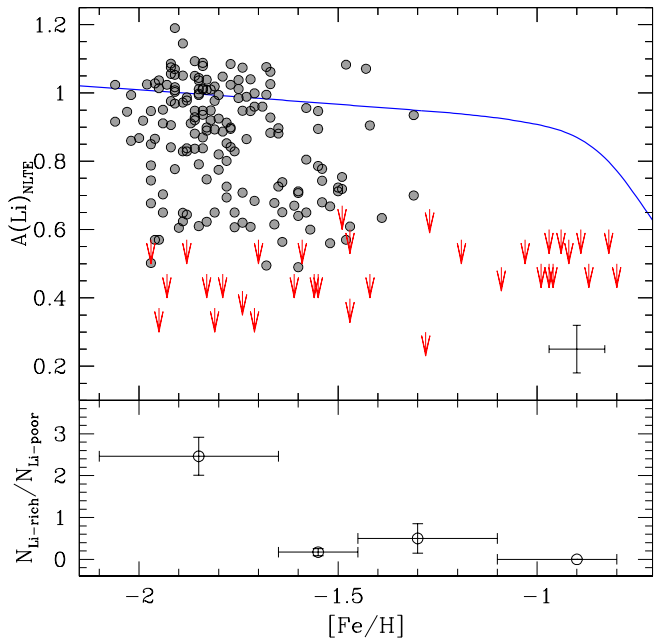}
\caption{Upper panel: behaviour of A(Li) as a function of [Fe/H] 
(grey points are the true measures, red arrows mark the upper limits). 
The blue line shows the theoretical behaviour of A(Li) as a function of 
[Fe/H] in LRGB stars starting from the same initial A(Li).
Lower panel: behaviour of the fraction of number of stars with A(Li)$>$0.8 
and with A(Li)$<$0.8 as a function of [Fe/H] computed for four metallicity bins.
Vertical error-bars represent the uncertainties in the number ratio, 
the horizontal error-bars denote the corresponding metallicity range.}
\label{li_dist}
\end{figure}

The upper panel of Fig.~\ref{li_dist} shows the behaviour of A(Li) as a function of [Fe/H]  
for the true measures (166 stars, grey circles) and the upper limits (33 stars, red arrows). 
Considering only the true measures, the A(Li) distribution ranges 
from 0.47 to 1.19 dex.
The A(Li) distribution clearly peaks at A(Li)$\sim$1 dex but with the presence of 
a significant fraction of stars with A(Li)$<$0.8 dex.
Stars with A(Li)$\sim$1 dex are found only at [Fe/H]$<$--1.3 dex, while Li-poor stars 
are found at all the metallicities.
In particular, the A(Li) distribution for stars with [Fe/H]$<$--1.6 dex 
(corresponding to the main population 
of Omega Centauri, RGB-MP) is dominated by stars with A(Li)$\sim$1 dex. 
In the [Fe/H] range between --1.6 and --1.3 dex (likely connected to RGB-Int1) 
the distribution is dominated by stars with A(Li)$<$0.8 dex, while the 
component at A(Li)$\sim$1 dex decreases in number.
For [Fe/H]$>$--1.3 dex the component with A(Li)$\sim$1 dex totally disappears and 
all stars have A(Li) lower than $\sim$0.6--0.7 dex.

The lower panel of Fig.~\ref{li_dist} shows the run of the ratio between the number 
of Li-rich and Li-poor stars (defined assuming A(Li)=~0.8 dex as threshold) 
as a function of the iron abundance. The data have been grouped in 4 metallicity bins, 
chosen according to the metallicity distribution (Fig.~\ref{irondis}),
centred at [Fe/H]=--1.85, --1.55, --1.30 and -0.9 dex, and with width of 
0.45, 0.20, 0.34 and 0.3 dex.
The stars with [Fe/H]$<$--1.65 dex show a constant number ratio that drops 
at higher metallicity, reaching 0 for the most metal-rich stars where 
all the stars have A(Li)$<$0.8.  

In order to assess whether the lack of Li-rich stars in the metal-rich regime 
is real or due to small number statistics, we performed a MonteCarlo simulation 
from a A(Li) distribution resembling that observed for stars with [Fe/H]$<$--1.6 dex. 
We extracted from this distribution 10,000 samples of 40 stars and 10,000 samples 
of 15 stars, corresponding to the number of stars observed in the third and forth metallicity 
bin, respectively. We found that the probability to observe a $N_{Li-rich}/N_{Li-poor}$ 
number ratio smaller than 0.5 is 0 and less than 0.03\% for the two samples,
ruling out that the observed lack of stars with A(Li)$\sim$1 dex is due to the 
small number of stars. Also, we checked that different choices for the metallicity bins 
do not change this conclusion.
Hence, the difference between the A(Li) distributions ofo metal-poor and 
metal-rich stars is real.

The peak of the number distribution of A(Li) for the metal-poor stars well agrees 
with the A(Li) abundances previously measured in Population II LRGB stars, i.e. 
in Milky Way field stars \citep{m12}, in Galactic GCs 
\citep{lind09,m11,dorazi14,dorazi15} and in the extra-galactic GC M54 \citep{muc14}.
This finding confirms the previous result (based on dwarf stars) by \citet{monaco10} 
that Omega Centauri formed with the same lithium content of other Population II stars, 
formed in the Milky Way.
Considering only the stars with [Fe/H]$<$--1.3 dex, the main peak of the A(Li) 
number distribution (A(Li)$>$0.8 dex) includes about 60\% of the stars. This fraction 
well matches the fraction of the true measures of A(Li) among dwarf 
stars of Omega Centauri analysed by \citet{monaco10} in a similar 
range of metallicity ([Fe/H]$<$--1.4 dex).

%%%%%%%%%%%%%%%%%%%%%%%%%%%%%%%%%%%%%  Li distribution

The most interesting feature in this A(Li) distribution is the lack of stars 
with A(Li)$\sim$1 dex at metallicities higher than [Fe/H]$\sim$--1.3 dex. 
The blue line shown in Fig.~\ref{li_dist} is the expected run of A(Li) with [Fe/H] 
in LRGB stars assuming that they are all born with the same initial A(Li), according 
to models from \citet{m12}  and assuming an age of 12.5 Gyr.
In particular, we assumed an initial value able to match 
the observed A(Li) of the peak of the number distribution of the metal-poor component;
however the shape of the curve is independent of the assumed value for the initial A(Li). 
The model predicts that A(Li) in LRGB stars mildly decreases over a large range 
of [Fe/H] and only for [Fe/H]$>$--0.9 dex should rapidly drop due to the deeper convective envelope 
of such metal-rich stars. 
If we assume that all the stars in Omega Centauri formed with the same initial Li content, 
a clear drop of the Li abundance is expected at [Fe/H]$>$--0.9/--0.8 dex. 
Instead, the stars of Omega Centauri with [Fe/H]$>$--1.3 dex 
are systematically depleted in lithium.
We recall that LRGB stars of GCs with [Fe/H]$\sim$--1.1/--1.2 dex 
\citep[see e.g. M4, \citet{m11,monaco12} and NGC~2808,][]{dorazi15}
have A(Li) compatible with that measured in metal-poor clusters.

We investigated if other parameters can affect the behaviour of A(Li) 
in these stars, in particular the helium mass fraction Y and 
the stellar age. Several photometric observations (like the extended blue horizontal branch 
and the splitting of the main sequence) reveal a large spread in the initial He content 
of Omega Centauri stars. 
In particular, the most metal-rich populations have been suggested to have a helium 
mass fraction up to Y$\sim$0.35 \citep{norris04,piotto05,sollima05b,renzini08}. 
We checked that for metallicity [M/H]=--1.01 dex, 
an increase of Y from 0.24 up to 0.35 changes the value of A(Li) in LRGB stars by 
only 0.03 dex. 

The precise age spread in Omega Centauri is still debated and the 
metal-rich populations are proposed to be coeval to the main population or 
significantly younger by some Gyr \citep[see e.g.][]{ferraro04,freyhammer05,villanova14,tailo16}.
However, a decrease of the age from 12.5 Gyr to 8.5 Gyr (both adopting Y=~0.25 and Y=~0.35) 
leads to a decrease of A(Li) by $\sim$0.05--0.06 dex with respect to 
the value calculated for 12.5 Gyr and Y=~0.25.
Therefore, an increase of the He content and/or a decrease of the age 
cannot explain the drop of A(Li) observed for [Fe/H]$>$--1.3 dex. 

Finally, we note that the model predictions for the surface A(Li) of the observed RGB stars 
(blue line in Figure~\ref{li_dist}) do include potential surface Li depletion during the MS, due to burning 
at the bottom of the convective envelopes.

\section{Li-Na anti-correlation}

\begin{figure}
\includegraphics[width=\columnwidth]{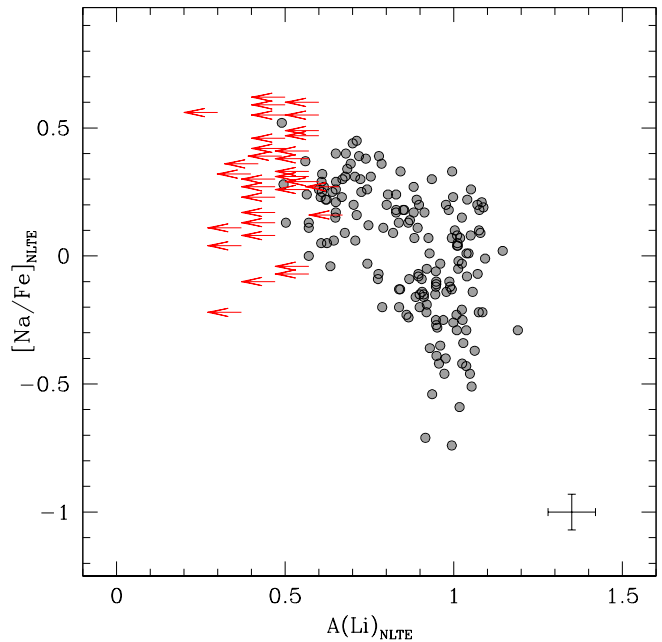}
\caption{Behaviour of [Na/Fe] as a function of  A(Li) for the entire spectroscopic sample 
studied in this work (grey circles). Red arrows mark the upper limits for A(Li). }
\label{lina}
\end{figure}

\begin{figure}
\includegraphics[width=\columnwidth]{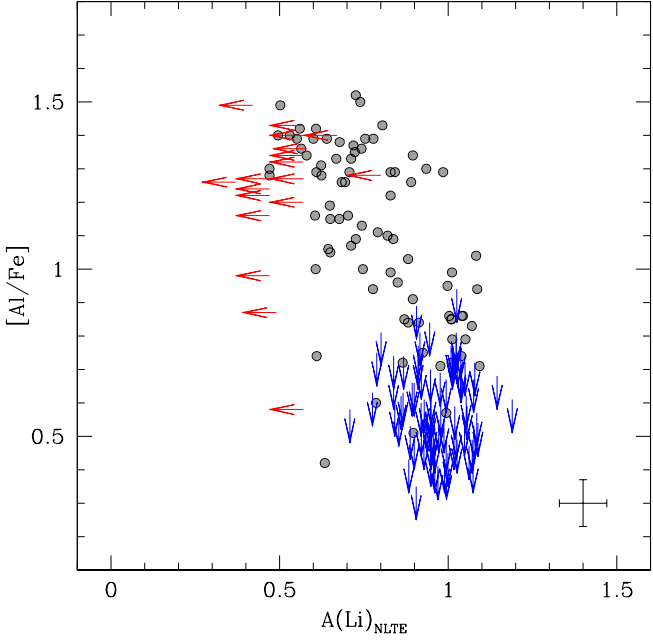}
\caption{Behaviour of [Al/Fe] as a function of  A(Li) for the entire spectroscopic sample 
studied in this work (grey circles). Red arrows mark the upper limits for A(Li), 
blue arrows the upper limits for Al.}
\label{lial}
\end{figure}

\begin{figure}
\includegraphics[width=\columnwidth]{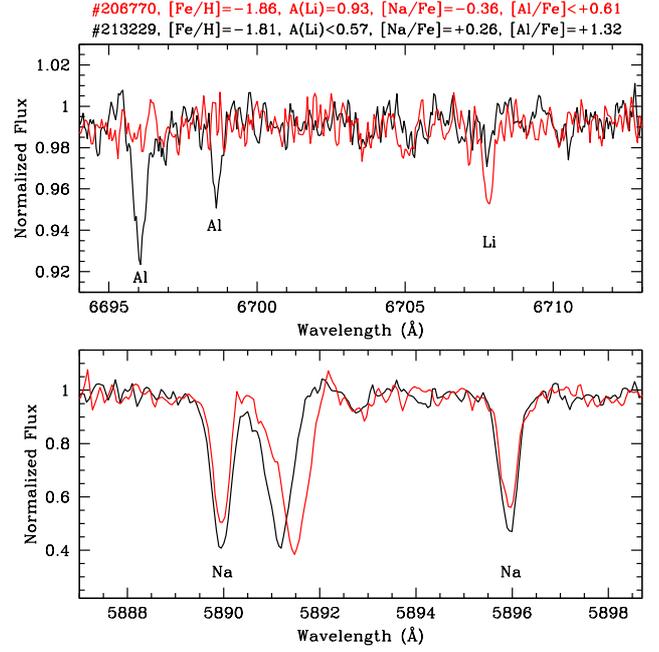}
\caption{Comparison between two stars of Omega Centauri, namely \#206770 and \#213229 
(red and black curves respectively), 
with similar atmospheric parameters and metallicity but different depths for 
Li, Na and Al lines. 
The two strong lines at $\sim$5891 \AA\ visible in the lower panel and shifted each other 
are Na interstellar lines.}
\label{spec1}
\end{figure}

The comparison between Li and Na abundances provides important clues to understand 
the nature of the Li-poor stars observed at all the metallicities in Omega Centauri.
This stellar system has a wide range of [Na/Fe], from [Na/Fe]$\sim$--0.5 dex to $\sim$+0.6 dex. 
Note that our [Na/Fe] distribution is shifted toward lower values with respect to those 
measured by \citet{johnson10} and \citet{marino11}, because these two works do not take into 
account departures from the local thermodynamical equilibrium (the corrections for the 
Na abundances are usually negative).
Fig.~\ref{lina} and Fig.~\ref{lial} show the run of [Na/Fe] and [Al/Fe] as a function of A(Li), 
respectively. Red arrows mark the upper limits for Li, blue arrows the upper limits for [Al/Fe]. 
In both plots a clear anti-correlation is found, with the stars with the higher A(Li)
covering a large range of [Na/Fe] and [Al/Fe], while the stars with A(Li) lower than 
$\sim$0.8 dex have systematically higher [Na/Fe] and [Al/Fe] abundances.
A direct evidence of this behaviour is shown in Fig.~\ref{spec1}, where 
the spectra of two stars stars with similar atmospheric parameters and metallicity 
are compared (namely \#206770 and \#213229, red and black curves respectively). 
As clearly visible, the spectrum of \#213229 shows Al and Na lines stronger 
than those of \#206770 (where the Al lines are totally lacking) but also an undetectable Li line, 
clearly visible in the other spectrum.

%%%%%%%%%%%%%%%%%%%%%%%%%%%%%%%%%%%%%  Na-Li

%%%%%%%%%%%%%%%%%%%%%%%%%%%%%%%%%%%%%  Al-Li

%%%%%%%%%%%%%%%%%%%%%%%%%%%%%%%%%%%%%  spec1

\begin{figure}
\includegraphics[width=\columnwidth]{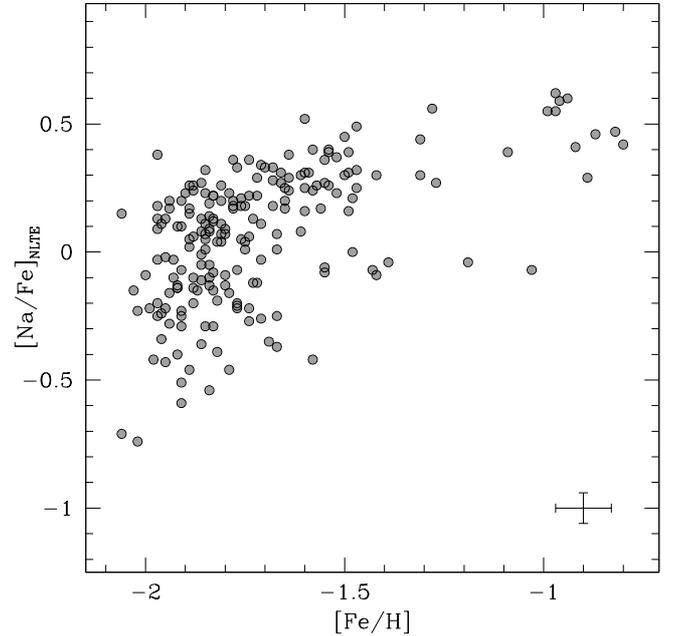}
\caption{Behaviour of [Na/Fe] as a function of [Fe/H].}
\label{nafe}
\end{figure}

The Li-Na anticorrelation is visible only in the metal-poor 
stars, because the presence of a significant star-to-star A(Li) scatter is 
restricted to [Fe/H]$<$--1.3 dex (see Fig.~\ref{li_dist}). 
In particular, the stars with low A(Li) are all significantly 
enriched in Na, even if other stars with similar high Na have normal A(Li). 
Fig.~\ref{nafe} shows the behaviour of [Na/Fe] as a function of [Fe/H] 
for our sample \citep[a similar run has been found also by ][]{johnson10,marino11}.
[Na/Fe] increases with the iron content and the higher 
Na abundances are measured in the most metal-rich stars. 
Hence, stars with [Fe/H]$>$--1.3 dex are all depleted in Li but 
enriched in Na.

\begin{figure}
\includegraphics[width=\columnwidth]{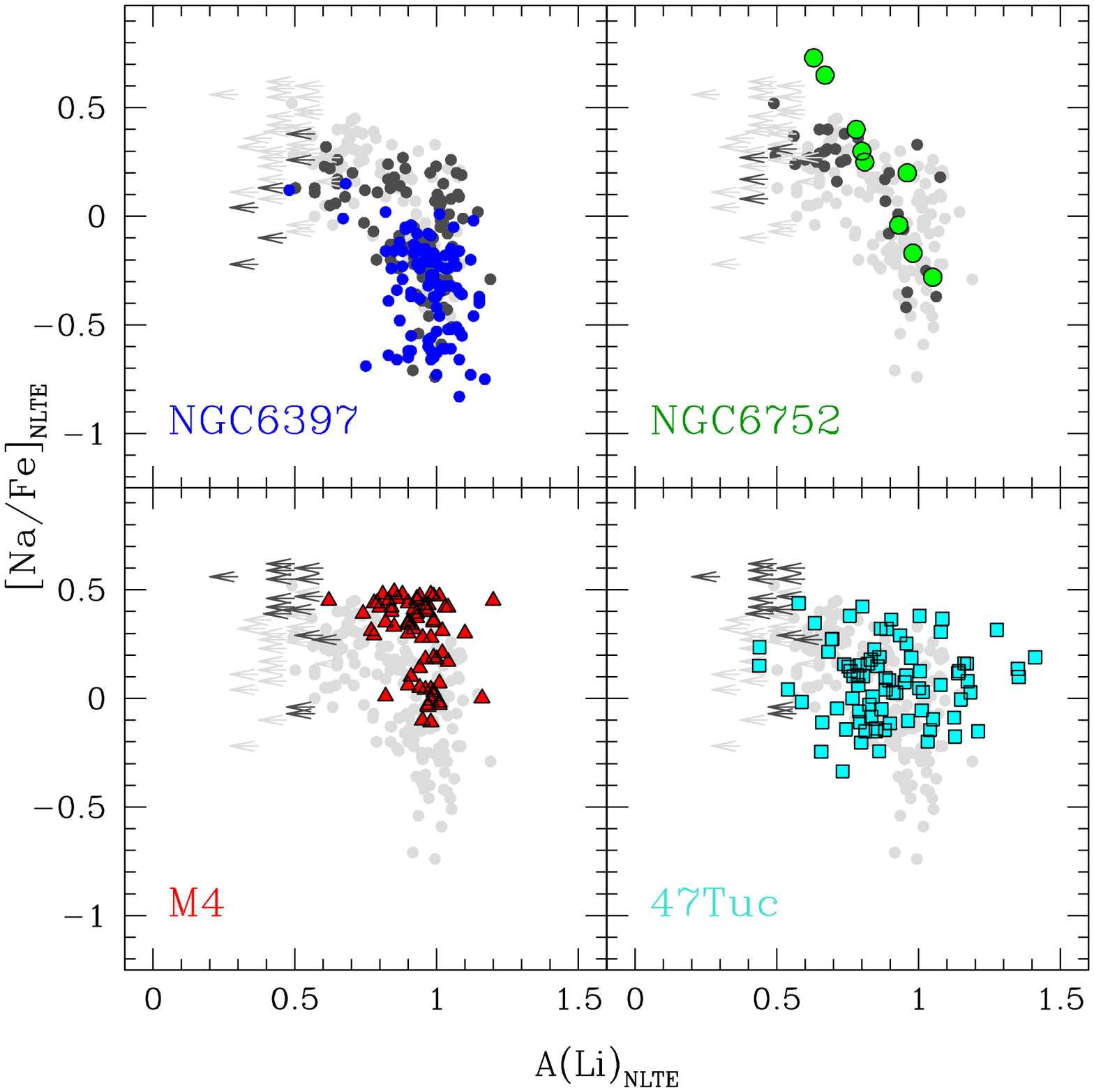}
\caption{Comparison between the [Na/Fe]-A(Li) distribution of the 
stars in Omega Centauri (grey points and arrows) and the abundances 
measured in dwarf stars of some
mono-metallic GCs from the literature, namely 
NGC~6397 \citep[blue circles,][]{lind09}, 
NGC~6752 \citep[green circles,][]{pasquini05}, 
M4 \citep[red triangles,][]{monaco12} and 47Tucanae \citep[cyan squares,][]{dobrov14}. 
The A(Li) abundances in these four 
clusters have been lowered in order to match their median value with the A(Li) measured in LRGB 
stars of Omega Centauri. Light grey points/arrows indicate the entire sample 
of stars in Omega Centauri studied here, dark grey points/arrows are the target stars 
selected with a metallicity similar to that of the reference Galactic GC shown in each 
panel.}
\label{compar}
\end{figure}

We compare the Li-Na distribution of Omega Centauri with those measured 
in mono-metallic GCs, namely NGC~6397, NGC~6752, M4 and 47 Tucanae. 
The distributions of Li-Na of these clusters are shown in Fig.~\ref{compar} 
in comparison with that of the LRGB stars in Omega Centauri: 
light grey points/arrows is the entire sample of target stars studied here, 
while dark grey points/arrows show the target stars of Omega Centauri 
selected with metallicities close to that of the comparison GCs included
in each panel.
Because the Li abundances in these GCs are measured in dwarf stars we shifted them 
in order to match the median A(Li) of each cluster with the peak of the A(Li) 
distribution of Omega Centauri (that is lower due to the 
effect of the FDU).

% NGC6397
In NGC~6397 (blue circles in Fig.~\ref{compar}) most of the 100 dwarf stars 
studied by \citet{lind09} have a similar A(Li) but over a large range of Na 
(hence corresponding to 1P and 2P stars). Among these 100 targets, three 
stars have A(Li) lower than the other stars by $\sim$0.4--0.5 dex and they 
have the highest values of [Na/Fe] of the sample.

% NGC6752
NGC~6752 (green circles) shows a clear Li-Na anticorrelation \citep{pasquini05} 
where the most Li-poor stars have A(Li) lower by $\sim$0.4 dex with respect 
to the most Li-rich stars. At variance with NGC~6397 (and also Omega Centauri) 
where most of the stars exhibit a large Na spread but similar A(Li), in NGC~6752 
the Na and Li abundances in the studied stars follow a linear behaviour. 
Even if the sample discussed by \citet{pasquini05} includes only 9 dwarf stars, 
its large spread in A(Li) has been confirmed by \citet{shen10} 
with a sample of 112 dwarf stars (and where a Li-O correlation has been detected).

% M4
\citet{monaco12} present Na and Li abundances in 70 dwarf stars in M4 (red triangles 
in Fig.~\ref{compar}), excluding one Li-rich star. 
The Na-Li distribution in M4 is similar to that of NGC~6397, with most of the stars 
with similar Li content and different [Na/Fe] but a larger star-to-star scatter in A(Li) 
among the 2P stars.
Note that the most Li-poor star in this sample has a A(Li) lower by $\sim$0.3 dex 
than the median A(Li) value and it is among the most Na-rich stars.

% 47Tuc
47Tucanae (cyan squares in Fig.~\ref{compar}) exhibits a large star-to-star scatter in its Li content. 
\citet{dorazi10tuc} and \citet{dobrov14} from the analysis of 110 dwarf stars in this cluster 
do not find clear evidence of Li-Na anticorrelation, while a hint of Li-O correlation 
is detected.

Other GCs have been investigated by \citet{dorazi14,dorazi15} that discuss the behaviour 
of A(Li) in LRGB stars as a function of Al abundance. 
In most of these GCs (namely NGC~362, NGC~1904, 
NGC~5904 and NGC~6218) all the stars have the same A(Li) regardless of the Al abundance. 
On the other hand, in NGC~2808 some Al-rich LRGB stars have A(Li) lower by $\sim$0.3 dex 
than that in other stars with similar Al content.

\section{Summary and conclusions}
We derived surface lithium abundances for 199  LRGB stars members of 
the stellar system Omega Centauri and distributed over its entire 
metallicity range. The main results are:
\begin{itemize}
\item the A(Li) number distribution of LRGB stars in Omega Centauri is peaked at A(Li)$\sim$1 dex, 
compatible with the abundances measured in LRGB stars belonging to other 
GCs and in metal poor field stars and this abundance can be 
considered as the normal A(Li) in Population II LRGB stars. 
Additionally, the A(Li) distribution of Omega Centauri
shows also a prominent Li-poor tail (A(Li)$\lesssim$0.8 dex);
\item the stars with normal A(Li) are found at [Fe/H]$<$--1.3 dex, 
while the Li-poor stars are found at all the metallicities.
All the stars with [Fe/H]$>$--1.3 dex exhibit low A(Li) values;
\item a clear Li-Na anticorrelation is found. The stars with normal A(Li) 
cover a large range of [Na/Fe],  with A(Li)$\lesssim$0.8 dex
are characterized by enhanced values of [Na/Fe]. Similarly, a  Li-Al 
anticorrelation is detected.
\end{itemize}

\begin{figure}
\includegraphics[width=0.5\textwidth]{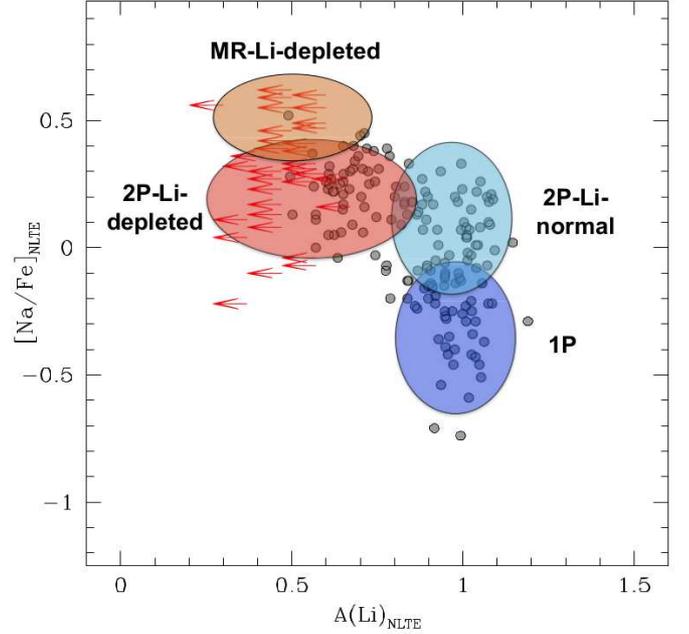}
\caption{Same of Fig.~\ref{lina} but with super-imposed the mean loci 
for the proposed classification of the Omega Centauri's stars according 
to their Li and Na abundances.}
\label{lipop}
\end{figure}

The A(Li) distribution in Omega Centauri is more complex 
than those observed in mono-metallic GCs, reflecting the peculiar (and not yet totally understood) 
chemical evolution of this system. 
We can draw a qualitative scheme to describe this distribution, 
classifying the stars according to their, Li, Na and Fe abundances. In particular, 
[Na/Fe] has been used to discriminate between 1P (low [Na/Fe]) and 2P (high [Na/Fe]) stars. 
We identify  four main groups of stars, sketched in Fig.~\ref{lipop}:
\begin{enumerate}
\item 1P stars ([Fe/H]$<$--1.3 dex), with low [Na/Fe] and normal Li. 
\item 2P stars ([Fe/H]$<$--1.3 dex) enriched in [Na/Fe] and with the same 
(normal) A(Li) measured in 1P stars; hereafter we refer to these stars 
as {\sl 2P-Li-normal}
\item 2P stars ([Fe/H]$<$--1.3 dex) enriched in [Na/Fe] like 
the {\sl 2P-Li-normal} stars but depleted Li (A(Li)$\lesssim$0.8 dex) 
and here named {\sl 2P-Li-depleted};
\item Stars with [Fe/H]$>$--1.3 dex (labelled as {\sl MR-Li-depleted}) 
that have A(Li)$\lesssim$0.6 dex and high [Na/Fe] abundances. In particular, 
the [Na/Fe] abundance of these stars is slightly higher than that of 
2P stars, because the [Na/Fe] increases with the metallicity 
\citep[see Fig.~\ref{nafe} and also][]{johnson10,marino11}.
\end{enumerate}

The classification proposed in Fig.~\ref{lipop} provides only a simple
guideline to highlight the complexity of the Li-Na distribution 
in this system, and the main features to be explained
 by models for the formation of Omega Centauri. It does not imply a 
specific formation timeline.

The overall picture is that mono-metallic GCs (see Fig.~\ref{compar}) show, to a lesser extent, 
the same sub-populations identified in the metal-poor stars of Omega Centauri.
These clusters are dominated by 1P and {\sl 2P-Li-normal} stars, with 
the presence of a minor component of Na-rich, Li-poor stars (even if 
the identification of a clear boundary between {\sl 2P-Li-normal} and {\sl 2P-Li-depleted} 
populations is not trivial). 
On the other hand, the fraction of {\sl 2P-Li-depleted} stars observed in Omega Centauri 
($\sim$30\%) turns out to be exceptional  with respect to the other GCs, where the 
fraction of these stars is about 3-5\%.

Within the canonical views of the formation of multiple populations,
1P stars would correspond to the first stellar generation
and we can consider their surface A(Li) as a good tracer of the initial
lithium content of the stellar system (after the effect of the FDU is
taken into account).
The high [Na/Fe] measured in the other (2P) stars indicates that these 
stars formed from a gas enriched by the products of proton-capture cycles 
(occurring at temperatures higher than that of Li-burning). 
The timeline of the formation of the 
2P stars (both with normal and depleted Li content) in Omega Centauri depends on the 
chemical evolution of the system and on the role played by the different 
polluter stars.  
The 2P-Li-depleted stars may have formed from a Li-poor or Li-free gas 
(diluted with pristine gas) coming from fast-rotating massive stars 
that are not able to produce fresh Li \citep{decressin07}. On the other hand, 
the high A(Li) measured among the {\sl 2P-Li-normal} stars suggests 
a production of new Li. 
Massive AGB stars are potentially able to produce Li through the
Cameron-Fowler mechanism
and they may explain the Li abundance in most of the Na-rich stars. 

However, theoretical models for A(Li) in GCs based on AGB stars as main polluters 
depend on several parameters, in particular the yields of the AGB models 
(the production of Li in these stars is sensitive to the stellar mass)
and the lithium abundance of the pristine material 
\citep[see][for a discussion on these assumptions]{dantona12}. 
\citet{dantona12} discussed for M4 the case of
polluters ejecta that are Li-free (a case similar to that of fast-rotating massive stars) 
and diluted with pristine gas,
finding only a mildly lower (by about 0.1 dex) A(Li) in 2P stars with respect to 1P stars.
Hence, high A(Li) in Na-rich stars could be explained also with polluter stars 
not able to produce fresh Li, depending on the details of the dilution process. 
The need of some fine-tuning to reproduce the uniform Li content observed 
in stars with a large spread in Na abundance, as well as uncertainties in the 
theoretical mass-loss rates and Li yields, reduce the predictive power of these models.
%%%%%

Another peculiarity of Omega Centauri is the low Li abundance in its metal-rich stars.
The stars with [Fe/H]$>$--1.3 dex exhibit an unexpected depletion of A(Li), 
not observed in GCs with comparable metallicities (i.e. M4 and NGC~2808) 
that show in most of their (dwarf or LRGB) stars A(Li) similar to that measured 
in metal-poor clusters.
We demonstrated that the effect of high metallicity, high He content and/or 
young age (that could characterize the high-metallicity stars in 
Omega Centauri) cannot explain the observed values, under the assumption 
that all the stars of Omega Centauri formed with the same initial A(Li). 
Hence, we conclude that the {\sl MR-Li-depleted} stars formed from a gas 
already depleted in A(Li) (similar to the {\sl 2P-Li-depleted} stars) and 
the observed Li depletion is not an effect of the FDU.
According to this finding, we should observe in the metal-rich main sequence 
stars of Omega Centauri a surface A(Li) smaller than the typical value of the 
{\sl Spite Plateau}. The observation of these stars is very challenging 
due to their faintness. \citet{monaco10} and \citet{pancino11b} provide upper limits 
for A(Li) in 18 stars belonging to the metal-rich sub-giant branch of Omega Centauri. 
Unfortunately, the upper limits for the initial Li content of these stars 
do not provide conclusive answers.

An additional hurdle with respect to mono-metallic GCs is the possible age spread of the stellar 
populations of Omega Centauri. There is no general consensus about the relative ages of the 
different stellar components of the system, in particular between the main (metal-poor) population and 
the most metal-rich one that could be virtually coeval 
\citep[see e.g.][]{tailo16} or several Gyr younger \citep[see e.g.][]{villanova14}. 
If the metallicity enrichment of Omega Centauri took place on short
timescales (in particular if the metal-rich stars are coeval or only
slightly younger than the other populations), the most metal-rich
stars have formed before other sources of Li production occur,
like novae and cosmic rays spallation \citep[see e.g.][]{romano99}.

Detailed theoretical models for the Li-Na behaviour in Omega Centauri 
are not available so far.
In particular, future theoretical models for Omega Centauri need to 
simultaneously explain:
(i)~the existence of stars with similar A(Li) but different [Na/Fe] 
(1P and {\sl 2P-Li-normal});
(ii)~the existence of stars with similar [Na/Fe] but different A(Li) 
(2P-LiR and {\sl 2P-Li-depleted});
(iii)~the {\sl MR-Li-depleted} component, with the lowest A(Li) and the highest [Na/Fe].

The discussed set of abundances will be crucial to put new constraints 
to the chemical enrichment history of this stellar system. \\

\begin{acknowledgements}
The authors thank the anonymous referee for his/her useful comments and suggestions. 
SV gratefully acknowledges the support provided by Fondecyt reg. n. 1170518.
LM acknowledges the support from "Proyecto interno" of the Universidad Andres Bello.
AM warmly thanks Federico Montaguti and Olma for their essential support during the last year.
\end{acknowledgements}

\bibliographystyle{apj}

\end{document}